# Sub-micrometer Nanostructure-based RGB Filters for CMOS Image Sensors

*Jonas Berzinš,[1,2,*] Stefan Fasold,[1] Thomas Pertsch,[1,3] Stefan M. B. Bäumer,[2] Frank Setzpfandt[1]*

[1]Institute of Applied Physics, Friedrich Schiller University Jena, Albert-Einstein-Str. 15, 07745 Jena, Germany

[2]TNO Optics Department, Stieltjesweg 1, 2628CK Delft, The Netherlands

[3]Fraunhofer Institute for Applied Optics and Precision Engineering, Albert-Einstein-Str. 7, 07745 Jena, Germany

*E-mail: jonas.berzins@uni-jena.de

**Abstract.** Digital color imaging relies on spectral filters on top of a pixelated sensor, such as a CMOS image sensor. An important parameter of imaging devices is their resolution, which depends on the size of the pixels. For many applications, a high resolution is desirable, consequently requiring small spectral filters. Dielectric nanostructures, due to their resonant behavior and its tunability, offer the possibility to be assembled into flexible and miniature spectral filters, which could potentially replace conventional pigmented and dye-based color filters. In this paper, we demonstrate the generation of transmissive structural colors based on uniform-height amorphous silicon nanostructures. We optimize the structures for the primary RGB colors and report the construction of sub-micrometer RGB filter arrays for a pixel size down to 0.5 μm.

In the past decades we have observed a rapid growth of the spatial resolution of color imaging devices[1–8]. This technological outburst was mainly driven by daily-use hand-held devices, such as mobile phones and compact optical cameras, but a high resolution also serves a role in industrial and environmental imaging[9]. The use of complementary metal-oxide-semiconductor (CMOS) image sensor with pigmented and dye-based color filters allowed to obtain pixels as small as 1 μm[1–3]. However, the further decrease of the pixel size is not a straightforward process. These filters require new materials and complicated multi-step fabrication[10]. In addition to that, the conventional filters rely on absorption, thus relatively high thicknesses are required[10] and, as the filters are arranged side-by-side, limitations such as crosstalk come into play[11,12]. An alternative is to use nanostructured surfaces, which due to their usually strong dispersion can be used to produce the so-called structural colors. The concept of structural colors is common in nature, e.g. blue *Morpho* butterfly wings are not pigmented, but are seen in color because the incident light is spectrally dispersed due to the nanostructured pattern on the wings[13,14]. This principle was quickly adapted in man-made structures using plasmonic nanoparticles[15–23]. However, the free-electron oscillations in metals are accompanied by a significant optical loss, which led to an investigation of high refractive index dielectric materials, such as silicon (Si), titanium dioxide ($TiO_2$), and gallium phosphate (GaP)[24]. Dielectric materials have lower intrinsic losses, thus can be used to obtain structural colors with a higher luminosity, which makes such nanostructures more suitable for applications, like color printing[25–33]. Furthermore, in contrast to plasmonics, dielectric nanostructures possess two types of fundamental resonances, electric dipole (ED) and magnetic dipole (MD) resonances[34–39]. This provides an additional degree of freedom in the tunability of the spectral response. The structural colors can be also achieved in transmission as the resonances provide dips in the spectra[40]. By varying the geometry of the dielectric nanostructures, the positions of the resonances can be tuned throughout the whole visible spectral range and beyond. This approach has been successfully used in producing a variety of transmissive colors including the realization of particular color filter arrays in RGB[12], CMY[41] and multi-spectral[42–45] arrangements.

In addition, an interesting aspect of nanostructured surfaces is the possibility to arrange them in a user-defined pattern with different local optical properties by changing just their lateral geometry. As the scattering of a single nanostructure can already provide a particular spectral response[35,46], a large density of different spectral functions is possible, which has been demonstrated in color printing with an extremely high number of dots per inch (DPI)[47]. Considering this, we tackle the question whether such high resolution could be also achieved in color filter arrays for imaging applications. Despite recent progress, the construction and analysis of a high-transmission nanostructure-based sub-micrometer filters, to our knowledge, has never been carried out. Here we will experimentally and numerically explore the limits of the size of the nanostructured RGB color filter arrays by carefully analyzing the crosstalk between different filters in order to prove the applicability of such filters in digital color imaging.

The structure we target in our work is schematically introduced in **Figure 1**(a). It consists of sub-micrometer RGB filters constituted of all-dielectric nanostructures, which are placed on top of a spacer layer on the CMOS image sensor. In order to realize different filter functions, the lateral geometry is varied. Amorphous Si being one of the most commonly used high refractive index and CMOS compatible materials was used for the nanostructures. An example of a realized single-size periodic array of nanostructures is presented in **Figure 1**(b). The inset of **Figure 1**(b) schematically depicts a unit-cell of such polarization independent disk-shape nanostructures, indicating the main tunable geometrical parameters: diameter, height, and period.

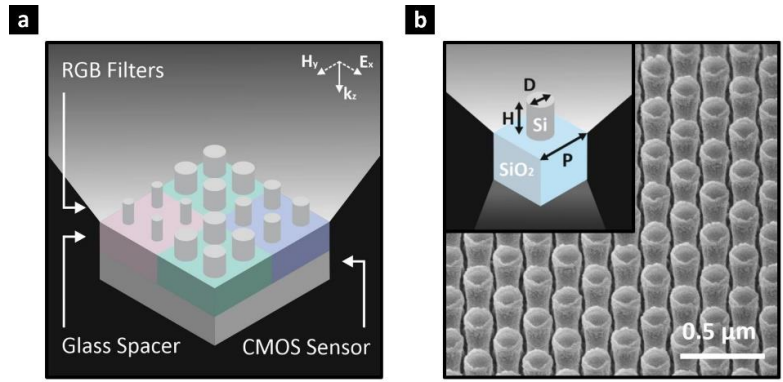

**Figure 1.** The proposed sub-micrometer dielectric nanostructure-based filters. **(a)** Graphic illustration of the implementation of sub-micrometer dielectric nanostructure-based RGB filters on top of a CMOS image sensor. **(b)** Scanning electron microscope (SEM) image of amorphous Si nanostructures on top of a glass substrate. The inset shows the main geometrical parameters of the filters: height (H), diameter (D) and period (P) of the nanodisks.

During the optimization of the filter functions, certain requirements coming from the system engineering and nanostructure fabrication have to be met. One important factor in the use of nanostructured filter arrays is the uniform-height to enable an efficient transverse patterning. To determine the optimal height for the simultaneous realization of all three primary RGB colors, we calculated transmission spectra for homogeneous periodic arrays of nanostructures varying in height, diameter, and period. In the optimization, also other conditions, e.g. the limited resolution of the electron-beam lithography[48], were taken into account. To compare the resulting colors, the transmission spectra were transformed into color codes by using CIE 1931 color matching functions, mimicking the human vision[49,50]. Then, utilizing the sRGB primary color basis, we obtained the color quality, a figure of merit introduced in previous work[40] (see Supporting Information S2 for details). The color quality $Q_\xi$, with $\xi \in \{R, G, B\}$ for red, green, blue, quantitatively compares the obtained color to the indexed color. It gains a maximum value of one, when the obtained color perfectly matches the primary color, and becomes zero when the achieved color is closer to a primary color different from the targeted one. Using this figure of merit, for each height the optimal realizations of R, G, and B filters were determined. As the most flawed part of the system limits its performance, we optimized towards the highest minimal value of the color quality between all three filters. We identified that a 175 nm height of the Si film gives the best compromise for realization of the colors. For this height, the optimal transverse dimensions of the nanostructures were found to be 80 nm, 105 nm and 150 nm in diameter, for R, G, and B respectively, and arranged in a period of 250 nm (see Supporting Information S3 for more details on parameters selection).

For experimental demonstration of color filters and confirmation of simulation results, we used a commercially available amorphous Si layer (see Supporting Information S4 for dispersion parameters) on top of a glass substrate, which was etched to the target height of 175 nm and later structured by electron-beam lithography and reactive ion etching (see Sample Fabrication and Characterization for more details). First, patches of $100 \times 100$ μm² size with nanostructures of different geometrical parameters were realized, which is a sufficient size to neglect boundary effects and observe the true color realized by the infinite nanostructure array. **Figure 2**(a) shows a matrix of color patches obtained from the measured spectra and arranged by their geometrical parameters. The colors obtained from the corresponding simulations are shown in **Figure 2**(b) and have a strong qualitative agreement. Due to it, the simulated spectra are further used to calculate the color quality. In **Figure 2**(c), the regions, for which $Q_R$, $Q_G$, or $Q_B$ are larger than zero, are indicated by the colored boundaries and provide additional knowledge about the limitations and tolerances of such filters. As shown in **Figure 2**(c), the R color filter seems to be relatively robust, while the B filters have a tolerance lower than 5 nm in case of a change in diameter. In contrast, the G filters are more tolerant, but are strongly affected by a change of the period. The color quality dependence on the period can be related to the near-field coupling of the neighboring nanostructures.

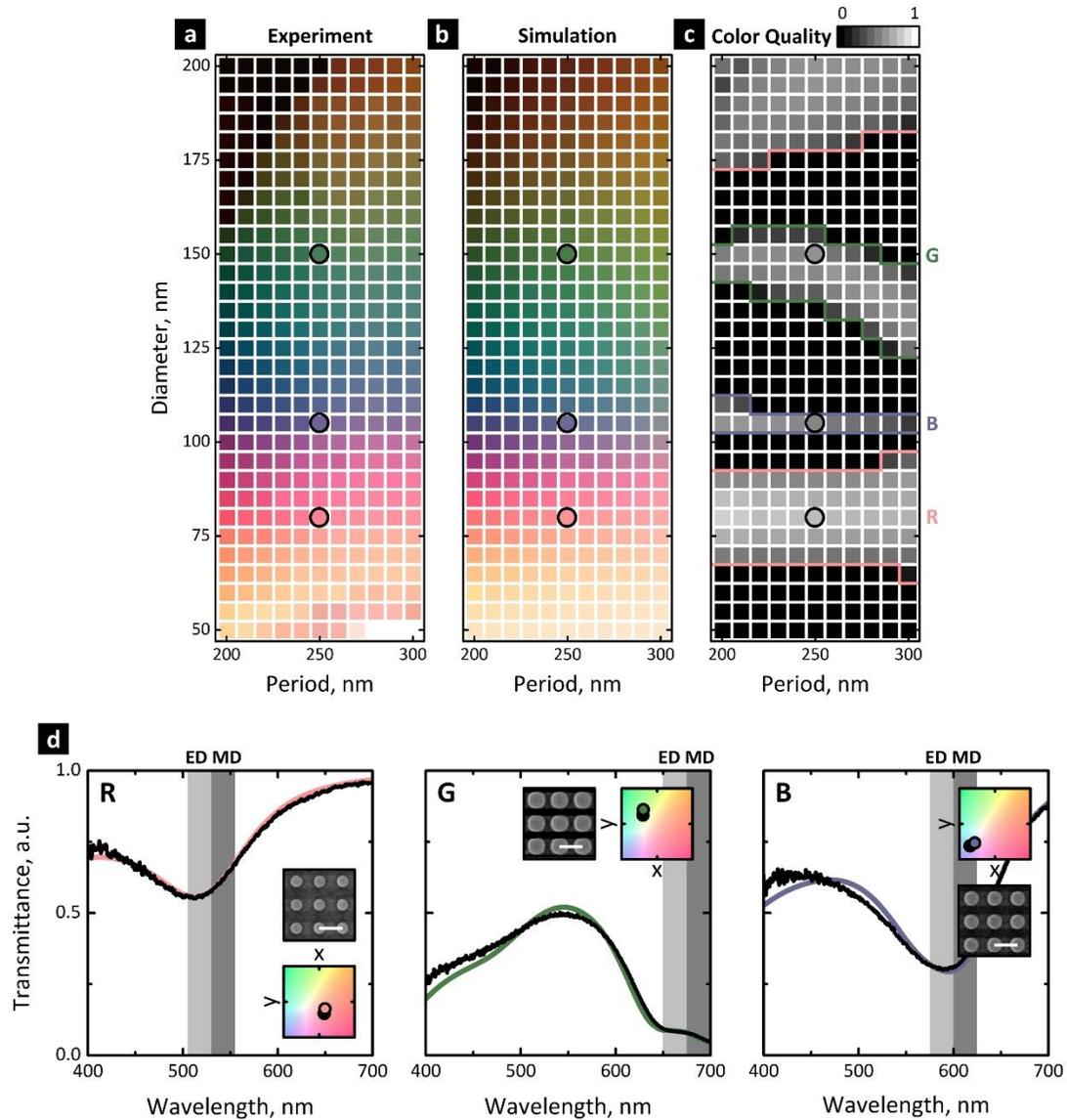

**Figure 2.** Visible spectral range filters using amorphous Si nanostructures. **(a)** Measured transmissive colors in dependence of period and diameter of the nanodisks with height of 175 nm. **(b)** Simulated colors, transformed from calculated spectra using CIE 1931 color matching functions and the sRGB primary color basis. **(c)** Color quality $Q_\xi$ calculated from the simulation data with the regions bigger than zero for R, G and B colors highlighted and the optimal RGB colors circled. **(d)** Measured (black line) and simulated (colored line) transmission spectra of optimal nanostructures for red (left), green (middle) and blue (right) color. The grey-shaded areas depict the central positions of the ED and MD resonances. The insets show the SEM images of the corresponding Si nanostructures (scale bar – 250nm) and the positions of the obtained colors in the chromaticity diagram.

The calculated and measured spectra of the optimal filters, denoted by the circles in **Figure 2**(a,b,c), are plotted in **Figure 2**(d). Experiment and simulation have excellent quantitative agreement. The small mismatches can be attributed to the fabrication tolerances and the possible formation of a Si oxide layer[26]. As expected and highlighted by the shaded areas, the central positions of the ED and MD resonances are distinguishable at the minima of the transmission function (see Supporting Information S1 for electric and magnetic field distribution analysis). As the spectral positions of the resonances depend on the size of the nanostructures[35,39], tuning their lateral geometry allows to select particular non-transmissive wavelengths through-out the full spectral range of the visible light (see Supporting Information S1 for a dataset plotted in the CIE 1931 chromaticity diagram), enabling the creation of the transmissive colors required for the RGB color filter arrays. The maximum transmittance of the optimal structures is approximately 90 % for the R filter at 650 nm, 50 % for G at 550 nm, and 60 % for B at 450 nm. The corresponding color quality values of the optimal filters are $Q_R = 0.74$, $Q_G = 0.59$, and $Q_B = 0.52$. In comparison, pigmented color filters can achieve slightly higher values of $Q_R = 0.94$, $Q_G = 0.82$, $Q_B = 0.76$[51]. However, considering the fabrication and scalability advantages of Si nanostructures as well as their insensitivity to strong illumination and non-degradation in case of ultraviolet light illumination, they are a viable alternative to established color filter technologies.

One of the main limitations of nanostructures is that their spectral response depends on the angle of incidence. To investigate the effect of different incidence angles on the color produced by the optimized RGB filters, we calculated their transmission spectra for angles up to 45°. In **Figure 3**(a) we show the resulting change in the transmission spectra. Whereas for the R filter only small changes can be observed, the transmission spectra of the other filters qualitatively change, with new maxima appearing and existing minima vanishing. These changes are due to the angle-dependent excitation cross-section of ED and MD resonances, their spectral shift and separation enabled by coupling of the nanostructures[52], and the excitation of higher order modes in the spectral range of interest. Even-though, we do not expect light scattering into higher diffraction orders at the normal incidence, due to the high effective refractive index of the filters they become apparent at the oblique incidence. To analyze the influence of all the effects on the color quality, we calculate $Q_\xi$ for the different incidence angles and plot it in **Figure 3**(b) for R, G, and B, respectively. As expected, the color quality decreases with increasing angle of incidence.

However, up to an angle of 25°, set by the B filter, color reconstruction can still be achieved and all filters produce distinguishable target colors. For the G filter, the maximum incidence angle is 40° and for the R filter it is larger than 45° (further analysis is given in Supporting Information S5).

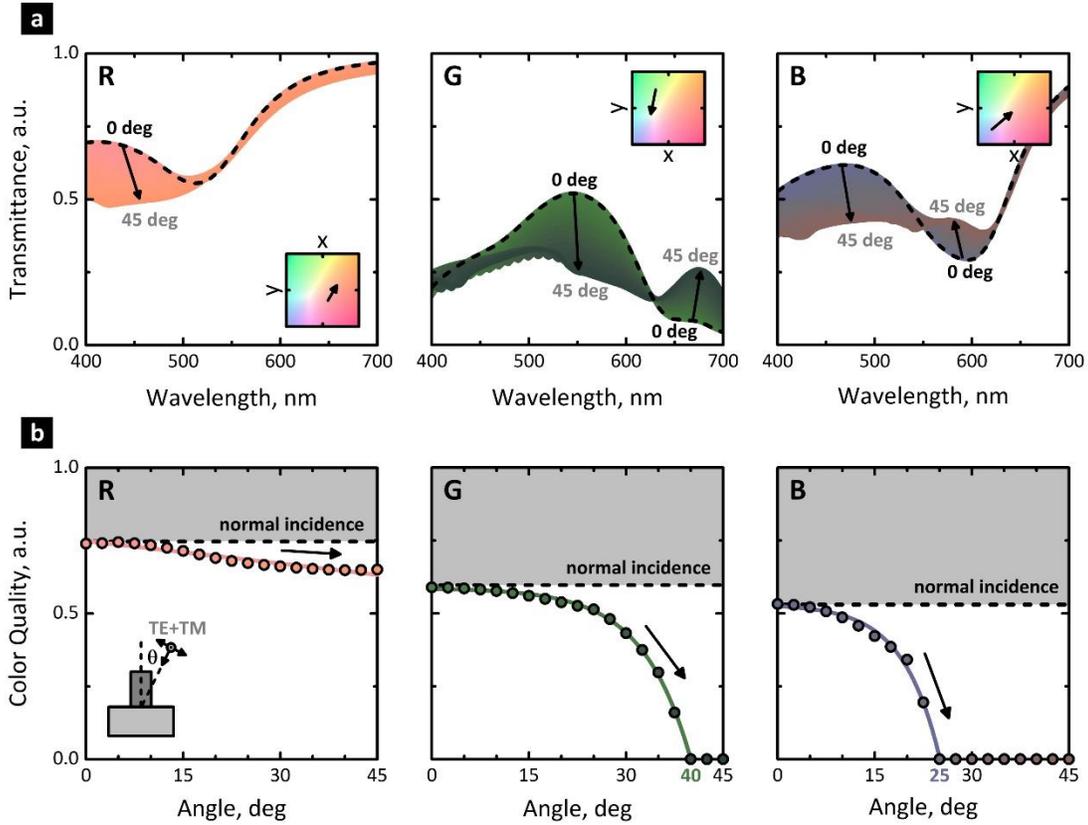

**Figure 3.** Angle tolerance analysis of optimized dielectric nanostructure-based RGB filters. **(a)** Transmission spectra of red (left), green (middle) and blue (right) filters and their change by the increase of the angle of incidence. The arrows highlight the transition from 0 degrees (highlighted by black dashed line) to 45 degrees in angle of incidence. The colors of the lines correspond to the transmitted colors. The chromaticity change is depicted in the chromaticity diagrams shown in the insets. The filters are illuminated by unpolarized light. **(b)** Color quality $Q_\xi$ change for red (left), green (middle) and blue (right) filters, compared to the normal incidence case (dashed line).

Finally, we study how the operation of the filters, arranged as small pixels in the Bayer-pattern [53], is influenced by the lateral size of the pixels. To this end, we consider the concept of a backside-illuminated CMOS image sensor[54], similar to the scheme depicted in **Figure 1**(a). Here we simplify the model by using a perfect absorber[5,55,56] at the interface of the photodiode, thus neglecting reflection and interference effects, meaning all the light passing through the filter array would be absorbed by the photodiode. For the geometric parameters of the nanostructures within the RGB pixels we used the optimum geometries determined earlier. The spacer layer with a refractive index of 1.46 between the filters and the photodiode was set to a thickness of 0.1 μm, which can be varied to a certain extent as shown in Supporting Information S6. We considered pixel sizes from 4 μm, equal to 16 periods of the nanostructure array, down to 0.25 μm, a single period. **Figure 4**(a) shows how the calculated transmission changes with the pixel size. For each wavelength, all 4 color pixels are illuminated and the transmission for each pixel is defined by the amount of light collected below that pixel and normalized to the fraction of the input power incident on that pixel. This means that also light scattered from the neighboring pixels will influence the measured transmission. In the transmission plots we notice that for smaller pixel sizes additional features appear, which are due to the coupling between the different nanostructures and scattering from one pixel to another. The biggest effect can be seen when working in the sub-micrometer scale. For instance, for 0.25 μm pixel size, sharp features appear at $\lambda \approx 500$ nm, corresponding to the period of the 2 x 2 filter array, while significant changes compared to infinite filters are also seen for longer wavelengths due to the coupling. To analyze the effect of the spectral reshaping on the colors, in **Figure 4**(b) we plot the color quality for the R, G, and B cases in dependence on the filter size. For 4 μm filter size, the calculated values are equal to the obtained in the case of infinite arrays: $Q_R = 0.74$, $Q_G = 0.59$, and $Q_B = 0.52$. For 0.5 μm filters positive values of $Q_R = 0.74$, $Q_G = 0.39$, and $Q_B = 0.36$ are maintained, meaning that the indexed colors can still be successfully discriminated. Hence, our analysis predicts that sub-micrometer color filters of 0.5 μm size, corresponding to just 2 x 2 nanostructures, would still be applicable. Unfortunately, with the selected R, G and B unit cells, going below 0.5 μm size is not possible due to the light diffraction and the coupling between the neighboring nanostructures. In correspondence to the spectral analysis, color qualities for 0.25 μm filter size drop significantly, with R and B diminishing to zero, meaning that the targeted colors are no longer distinguishable.

It should be noted that in large-scale color filter arrays multiple ways can be employed to reduce the crosstalk. The use of a micro-lens array on top of the filters could focus the incoming light, thus limiting the surface area of interaction[57,58]. A similar effect can be achieved by adding a mask layer acting as spatial filter, though this is connected to reduction of the effective sensor area. Also, even-though this limits the spatial resolution, the crosstalk can be reduced by separating the pixels by a gap between them, as allowed by the use of CMOS. However, for the sub-micrometer sized pixels the use of additional optical components or geometrical manipulation is very complicated and is outside of the scope of this work.

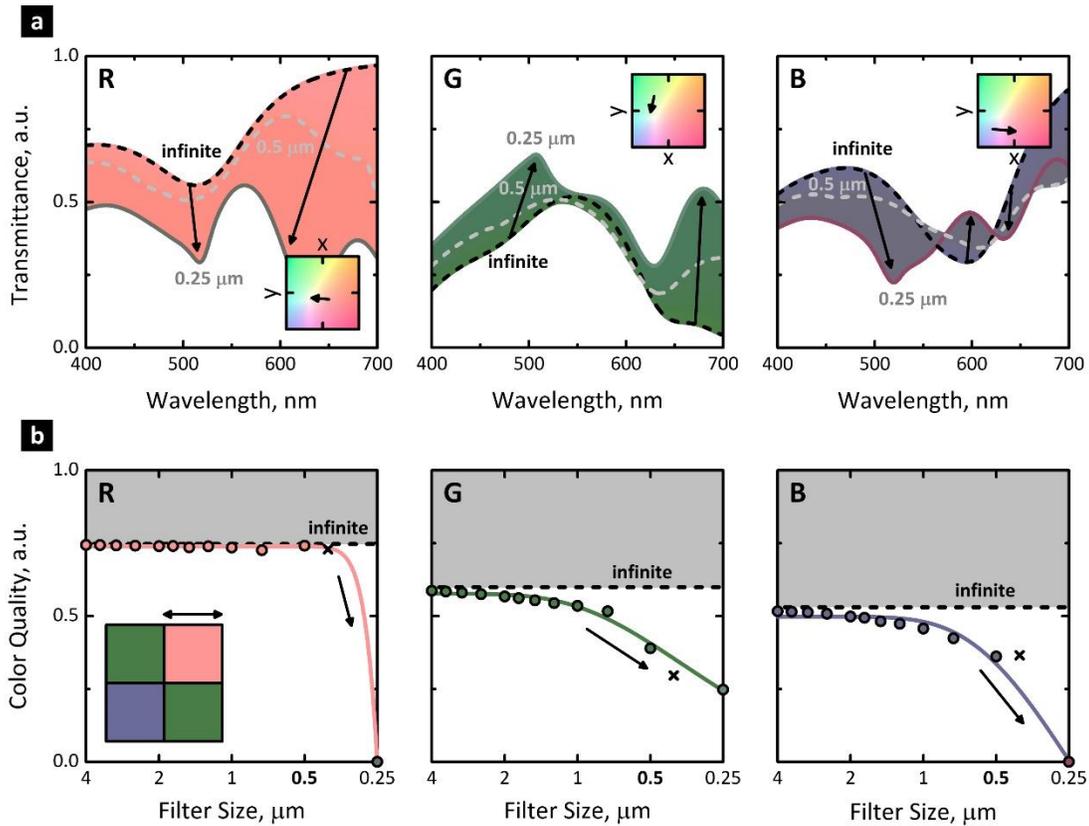

**Figure 4.** RGB filter array spectral response and crosstalk analysis based on color quality $Q\xi$. **(a)** Transmission of red (left), green (middle) and blue (right) filters, measured 0.1 μm below of all three color filters. Arrows, as well as the color coordinate change in CIE chromaticity diagram in the insets, show the spectral change by decrease of the pixel size. The infinite array case is highlighted by the black dashed line, the grey dashed line depicts the transmission functions of the 0.5 μm filters. (b) Color quality $Q\xi$ in dependence on the pixel size for red (left), green (middle) and blue (right) filters keeping the initial geometrical parameter. The color quality of 0.4 μm filters with tuned transversal parameters are depicted by a cross.

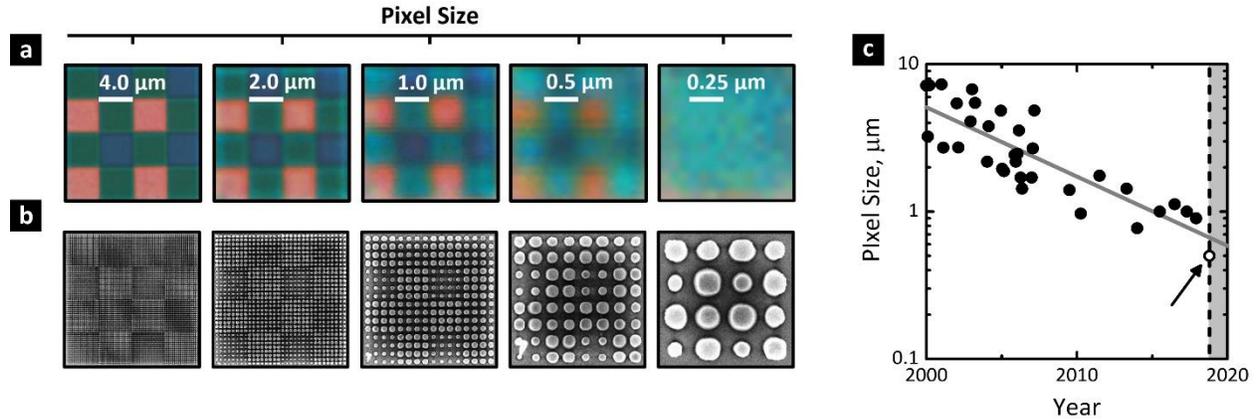

**Figure 5.** Down-scaling of RGB color filters assembled in Bayer pattern. **(a)** Optical microscope images of the 4 μm, 2 μm, 1 μm, 0.5 μm and 0.25 μm filters, imaged with 100x (NA = 0.9) objective. Scales are shown in the images. **(b)** Scanning electron microscope (SEM) images of the filters, same scales as in (a) are maintained. **(c)** The evolution of the pixel size in the last two decades and the position of pixels using 0.5 μm filters. Adapted data until the year of 2008[3] and extracted new points[2,4,6–8,12,16,21] shown in black dots till the date of submission (black dashed line), only filters arranged in color filter arrays are considered. The current result is highlighted by the white dot.

To experimentally verify the numerical predictions, we fabricated color filter arrays in Bayer pattern, consisting of two G, one R and one B pixels each. The pattern was made with additional pixels on the sides, thus forming 4 x 4 color filter arrays, as the fabrication procedure results into an over-etching of the structures at the edge of an array. The same pattern was repeated by reducing the size of the filters from 4 μm by a factor of two until a color pixel of single nanostructure is reached, resulting in the pixel sizes of 4 μm, 2 μm, 1 μm, 0.5 μm, and 0.25 μm. We characterized them using a transmissive white-light microscope. The obtained images, taken with a high numerical aperture (NA = 0.9) objective, are shown in **Figure 5**(a), together with SEM images of the structures in **Figure 5**(b). It is clearly visible, that the R, G, and B colors can be discriminated for pixel sizes down to 0.5 μm, which confirms the results of the numerical simulations. For 0.25 μm pixel size, the different filters cannot be resolved due to the diffraction limit[59], no conclusions about the colors can be drawn. Nevertheless, our results show that the RGB color filters as small as 0.5 μm size can be realized using Si nanostructures. The successfully obtained 0.5 μm filters are significantly smaller than previously reported filters in color filter arrays, as shown in **Figure 5**(c), where we compare our structures to other results presented in the last two decades[2–4,6–8,12,16,21]. Also, even-though at this point it was limited by the fabrication, theoretically the filter size smaller than 0.5 μm is possible by a denser distribution of the 2 x 2 nanostructures, as shown in **Figure 4**(b) with a positive color quality for all the colors produced by 0.4 μm filters. However, this requires re-optimization of the transversal geometry, e.g. decrease of the period, from 250 to 200 nm, for all of the filters and a slight change in the diameter of nanostructures composing the G filter, from 150 nm to 160 nm.

In summary, we have numerically and experimentally demonstrated sub-micrometer dielectric nanostructure-based RGB filters with high angle tolerance. Although amorphous Si is not the ideal material for such filters due to the broadband absorption losses limiting the maximum transmission in the short wavelength range, the realized filters show colors comparable to the conventional filters. In addition, they provide many advantageous properties, e.g. their robustness to high-intensity illumination and well-developed CMOS compatible fabrication process allowing their direct incorporation on top of backside-illuminated CMOS image sensors. Considering the achieved state-of-the-art 0.5 μm filters and the potential of them to be reduced even further by improving fabrication techniques, forming rectangular-shape nanostructures and distributing in a denser arrangement, the nanostructures present an intriguing opportunity for increasing the imaging resolution.

Furthermore, while the reported results focus on the primary RGB colors, a good control of transmissive colors is shown via the tunability of the spectral positions of ED and MD resonances. This potentially allows also the realization of multispectral filter arrays, which could make the full use of the enhanced camera resolution enabled by the sub-micrometer pixels and give an access to hitherto hardly obtainable information. The dielectric nanostructure-based filters have an undeniable potential and, we believe, their consequent use can greatly improve the technologies of digital color imaging.

**Numerical Simulations.** The simulations were carried out by using a commercial simulation tool based on the finite-difference time-domain method (FDTD Solutions, Lumerical Inc.). The initial optimization of the dielectric nanostructure-based filters was done using a 3D model of an infinite array of amorphous Si nanostructures, with wavelength dependent dispersion parameters given in Supporting Information S4, placed on top of the glass substrate, considering air as the covering medium. The model implemented periodic boundary conditions (PBC) on the sides and perfectly matched layers (PML) on the top and the bottom of the simulation domain. The mesh accuracy was set at the range of $\lambda/22$. Frequency domain field and power monitors were placed to record the electric and magnetic fields as well as the transmission and the reflection of the nanostructures, which is done through an integration of the power density over the surface of the monitor as a function of the wavelength. In the general case, the structure was excited by a normal-incidence plane wave source with a spectrum covering the visible spectral range, while the angle dependence simulations were done using a broadband fixed angle source technique (BFAST) to ensure a constant angle for all of the wavelengths within the range of interest. The simulations of the color filter array were done using the same PBC and PML boundaries, but extending the simulation domain to a pixel array of 2 x 2 pixels[57]. The simulation considered a semi-infinite substrate. For monitoring of the incident power separate monitors were put in a distance of 0.1 μm of all three different color pixels illuminated by normally incident plane-wave source.

**Sample Fabrication and Characterization.** The amorphous Si nanostructures were made from a commercial amorphous Si layer (Tafelmaier Dünnschicht-Technik GmbH), which has a height of 500 nm on top of a glass substrate. The Si layer was etched to the target height of 175 nm by argon (Ar) ion beam etching (Oxford Ionfab 300, Oxford Instruments) in steps. After each step the thickness of the sample was measured by an optical transmission setup and compared of the results with simulation data. Then a 30nm chromium (Cr) layer was deposited by ion beam deposition (Oxford Ionfab 300, Oxford Instruments). After spin coating of 100nm electron beam resist (EN038, Tokyo Ohka Kogyo Co., Ltd.) the sample was exposed by a variable-shaped electron-beam lithography system (Vistec SB 350, Vistec Electron Beam GmbH). The resist was developed for 30 s at room temperature and the created mask was transferred in Cr layer by ion beam etching (Oxford Ionfab 300, Oxford Instruments). Finally, the Cr mask was transferred in the Si layer by inductively coupled plasma reactive ion etching (Sentech SI-500 C, Sentech Instruments GmbH).

The reactive gas was tetrafluormethane (CF4). After etching the remaining resist and Cr mask was removed by acetone and a Cr etchant (ceric ammonium nitrate). The fabricated sample was analyzed visually by means of a white-light optical microscope (Axio Imager 2, Carl Zeiss AG) with an installed 5-megapixel camera (AxioCam MRc 5, Carl Zeiss AG). The white-light adjusted optical images were taken using a 100x magnification high numerical aperture (NA = 0.9) objective (EC Epiplan-Neofluar 100x/0.90 DIC M27, Zeiss GmbH). The scanning electron microscope (SEM) images were taken with a voltage of 3 kV (Helios NanoLab G3 UC, FEI Co.). The spectral analysis was done with a plane-wave illumination using an inverted optical microscope system (Axio Observer D1, Carl Zeiss AG) with integrated broad-band VIS/IR imaging spectrometer (iHR320, HORIBA Jobin Yvon GmbH).

**Supporting Information.** The supporting information includes: analysis of the spectral response, short introduction to color science and figure of merit, parameter optimization algorithm, dispersion parameters of amorphous silicon, angle tolerance analysis for polarized light and spacer thickness influence on color quality. The supporting information is given below.

**Acknowledgement.** This project has received funding from the European Union's Horizon 2020 research and innovation programme under the Marie Sklodowska-Curie grant agreement No. 675745 and German Federal Ministry of Education and Research (FKZ 03ZZ0434, FKZ 03Z1H534). The authors acknowledge Fabrizio Silvestri, Giampiero Gerini from TNO Optics Department and Isabelle Staude from Institute of Applied Physics, Friedrich Schiller University Jena for helpful discussions, also Dennis Arslan, Michael Steinert, Waltraud Gräf, Holger Schmidt, Thomas Käsebier and Jörg Fuchs from Institute of Applied Physics, Friedrich Schiller University Jena for technical assistance.

**Author Contributions.** F.S, T.P. and S.M.B.B. initiated the idea of using dielectric nanostructures for spectral filtering in the application of optical cameras, J.B. suggested the concept of the sub-micrometer pixels, F.S, T.P. and S.M.B.B. supervised the project. J.B. performed the finite-difference time-domain simulations, introduced the color optimization algorithm and optimized the RGB filters, S.F. designed and carried out the fabrication. J.B. characterized the sample and wrote the manuscript. All authors analyzed the data, read and corrected the manuscript before submission.

Supporting Information for

# Sub-micrometer Nanostructure-based RGB Filters for CMOS Image Sensors

*Jonas Berzinš,[1,2,*] Stefan Fasold,[1] Thomas Pertsch,[1,3] Stefan M. B. Bäumer,[2] Frank Setzpfandt[1]*

[1]Institute of Applied Physics, Friedrich Schiller University Jena, Albert-Einstein-Str. 15, 07745 Jena, Germany

[2]TNO Optics Department, Stieltjesweg 1, 2628CK Delft, The Netherlands

[3]Fraunhofer Institute for Applied Optics and Precision Engineering, Albert-Einstein-Str. 7, 07745 Jena, Germany

*E-mail: jonas.berzins@uni-jena.de

**S1. Analysis and Tunability of Spectral Response**

In order to explain an underlying mechanism of the spectral behavior of the dielectric nanostructure array, we show transmission, reflection and absorption of an array of nanodisk elements, see **Figure S1.1**(a). The chosen structures are of 80 nm in diameter with a period of 250 nm and a height of 175 nm. In addition, we observe the dynamics of the electric (E) and magnetic (H) fields. We select the wavelengths at the positions of the highest E and H intensities in the middle of the nanostructure, which happen to be at 520 nm and 540 nm, for E and H field respectively, as shown in **Figure S1.1**(b). We plot the E and H fields at the horizontal and vertical cross-sections of the nanostructure. As can be seen in **Figure S1.1**(c,d), the E field lines at 520 nm wavelength relate to an electric dipole (ED), while in **Figure S1.1**(e,f) the H field lines at 540 nm wavelength depict a magnetic dipole (MD). The excitation of ED resonance requires a collective polarization induced by the E field of the incident light, while the excitation of the MD driven by the E field of light that couples to displacement current loops[1]. This displacement current loop (as can be seen in **Figure S1.1**(f)) induces a MD moment, oriented perpendicularly to the E field polarization.

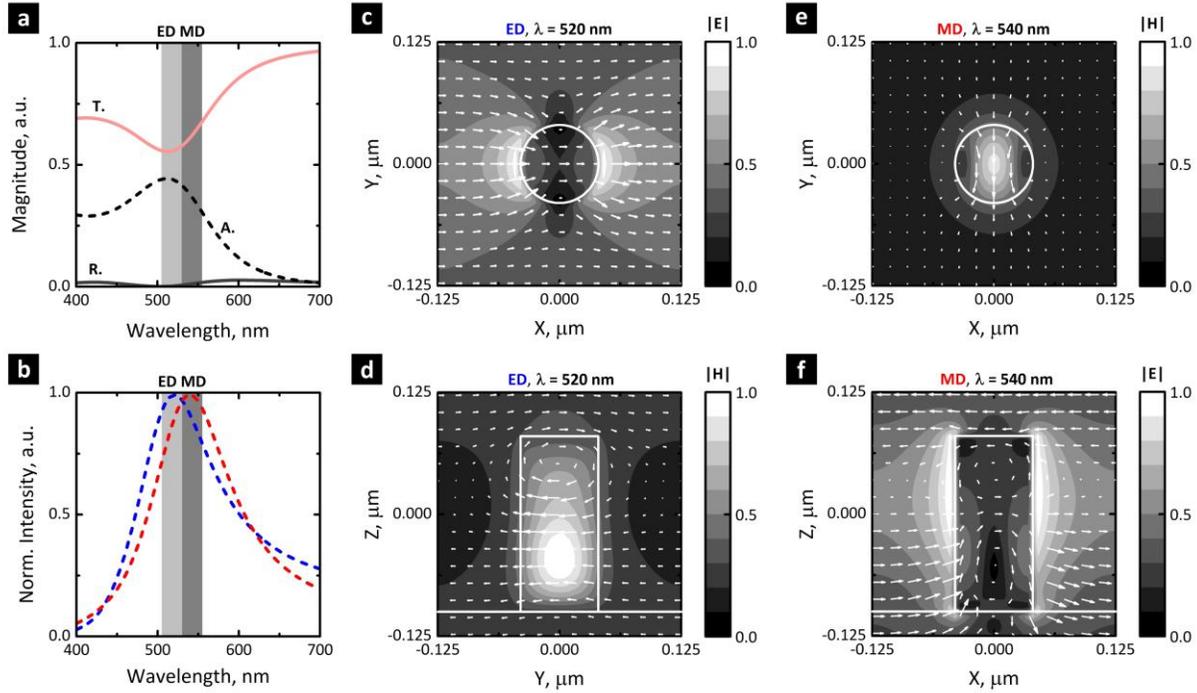

**Figure S1.1.** Spectral response analysis based on ED and MD resonances. **(a)** Magnitudes of transmission, reflection and absorption as well as highlighted wavelengths corresponding to positions of ED and MD resonances of a nanodisk of 80 nm diameter, 250 nm period and 175 nm height. **(b)** Specified normalized intensities of magnetic and electric fields inside the middle of the nanostructure. **(c)** E field at the horizontal and **(d)** H field at vertical cross-sections of the nanostructure associated to ED. **(e)** H field at the horizontal and **(f)** E field at vertical cross-sections of the nanostructure associated to MD resonance.

Based on Mie theory for spherical particles, the magnetic dipole resonance takes place when the wavelength of light inside the particle is comparable to its optical size[1,2]. However, for non-spherical nanostructures, such as in our case, a full analytical solution is not available. Despite that, the positions of the lowest order ED and MD can be estimated as they scale with the dimensions, such as diameter, as shown in **Figure S1.2**(a,b) for both, simulation and experiment. By increasing the size, also higher order modes appear in the range of interest. The scaling of the nanostructures is mainly limited by the geometrical fabrication possibilities, e.g. the smallest manufacturable size as well as the period between structures, comprising minimum and maximum possible size of the structure. Despite these limits, by optimizing array of nanostructures one may obtain conditions which allows to select particular bands of interest throughout the whole visible spectral range. In **Figure 1.2**(a,b) the diameter is varied from 50 nm to 200 nm, while having the period set to 250 nm and height set to 175 nm. The spectral representation in chromaticity diagrams, **Figure 1.2**(c,d), was carried out by the steps described in S2 section.

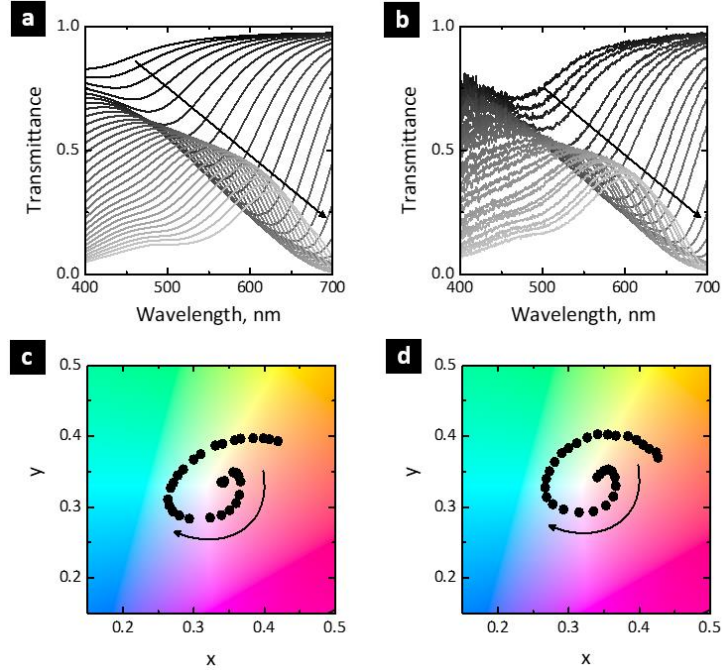

**Figure S1.2.** Spectral tunability by increase of nanostructure diameter. **(a)** Simulation. **(b)** Experiment. **(c,d)** Simulation and experimental data transformation into CIE 1931 chromaticity diagram. Height fixed at 175 nm, period set at 250 nm.

## S2. Color Science and Figure of Merit for Color Quality

In this section we shortly introduce the steps made for the estimation of a color quality, which is a derived merit for the search of the best filter functions. There are multiple ways to optimize RGB filter functions, such as optimization towards particular wavelength ranges[3], optimization in comparison to reference filters[4] or optimization according to CIE 1931 color matching functions[5]. The latter was used for all of the experiments and simulations in this work, and will be shortly discussed here.

The spectral composition of visible light can be inspected by the introduction of color matching functions, which are the base of the CIE Standard Colorimetric System[6]. Each of the color matching functions are linearly independent and are used to obtain tristimulus values (X, Y, Z) from the given spectral data. To calculate these values, we multiply our transmission spectra by a normalized power distribution of a day-light, then integrate over the specified color matching functions, as presented in the first equation. This allows us to numerically identify a color from a measured or simulated spectra.

$$X = \int T(\lambda) \times P_{D65}(\lambda) \times S_x(\lambda) d\lambda ,$$
$$Y = \int T(\lambda) \times P_{D65}(\lambda) \times S_y(\lambda) d\lambda , \qquad (1)$$
$$Z = \int T(\lambda) \times P_{D65}(\lambda) \times S_z(\lambda) d\lambda .$$

The chromaticity diagram is used to evaluate a color gamut. However, the chromaticity diagram does not show the value of luminance, thus full determination of the color requires a transformation into a particular color space. We suggest the use of HSV color space, which can be obtained by transforming tristimulus values to RGB and then the achieved values to HSV. The latter color space is further used for the formation of a figure of merit, color quality, which is adapted from [5], but slightly modified to follow a linear trend:

$$Q_\xi = \sqrt[3]{HSV_\xi} = \begin{cases} 0, & if\ \Delta H_\xi \geq \frac{\pi}{6} \\ \sqrt[3]{\left(1 - \frac{6}{\pi} \times \Delta H_\xi\right) \times S \times V} \end{cases}, \qquad (4)$$

where color is indexed by ξ, $\Delta H_\xi = |H_\xi - H|$ is a difference of the hue from the hue of a targeted color in radians, $S$ is the saturation and $V$ is the value of a color or basically the amplitude of the signal. This figure of merit results in only one positive $\xi_Q$ value, assigning the generated color to the closest primary color.

## S3. Parameter Cube and Optimization Algorithm

Considering that the material and the environment of the object is set, the amorphous silicon nanodisk array has three main variable parameters: height and diameter of the nanostructures as well as period between them. These parameters can be visually presented as a parameter cube (see **Figure S3.1**). While tuning the parameters we monitor the transmission, which is later transformed into a color and into a value of color quality as described in S2 section. In the algorithm we are looking for the maximum quality of the separate colors, but in order to limit our parameter space, e.g. select particular height, the key aspect to obtain the highest minimum between the RGB colors, as shown in **Figure S3.2**, in order to ensure that the quality of the worst filter would be as high as possible, securing the best values over all filters.

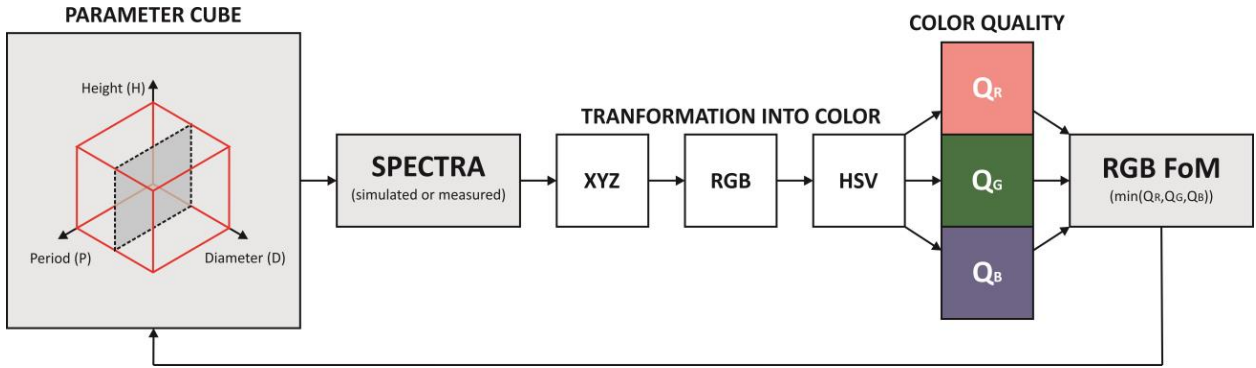

**Figure S3.1.** Schematic representation of parameter selection algorithm. From parameter sweep to color quality via spectral transformation into color.

In **Figure S3.2** we define a matrix of heights and periods and in each of the matrix points we sweep the diameter, starting from 50 nm to a value of 50 nm less than a period at the steps of 5 nm. We observe the change of the figure of merit regarding the height and period and can notice that the optimal values in case of R filter are almost constant, the values of G filter color quality vary at the boundaries of the selected range of variables, but in case of B we indicate a high fluctuation of the value by the change of height. This results into two peaks at 175 nm height and at 225 nm height, as seen in **Figure S3.2**(b), but at the former we notice a smaller deviation by the change of period, thus more tolerant in case of fabrication. At the selected height of 175 nm, the values between 210 nm period and 250 nm period are relatively similar, so the selection of the 250 nm period for the downscaling is justified by the minimization of the neighboring structures coupling in case of small size pixels as well as the equal division of 1 µm pixel into 4 x 4 array.

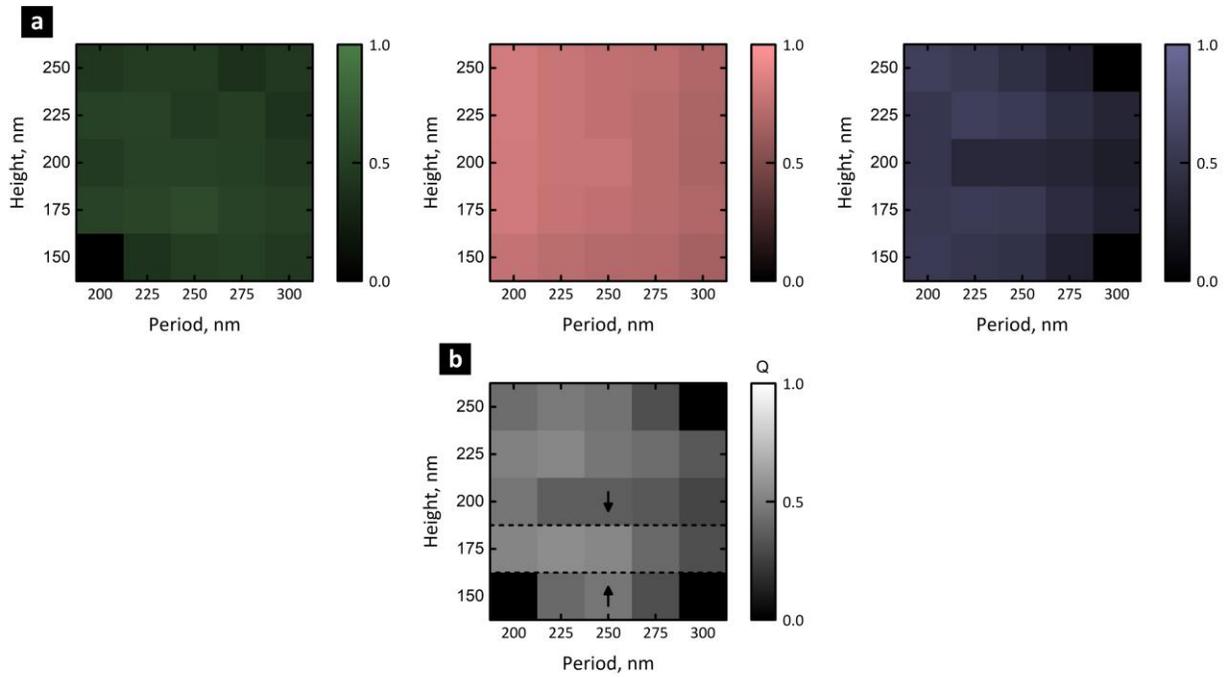

**Figure S3.2.** Geometrical height and period selection. **(a)** Highest RGB color qualities values in the matrix of height and period, after sweeping the diameter of the nanostructure. **(b)** Highest minimum values of RGB as the function of height and period. The selected height is 175 nm.

## S4. Dispersion Parameters of Amorphous Silicon

Amorphous silicon (Si) nanostructures were made from a commercial amorphous Si layer (Tafelmaier Dünnschicht-Technik GmbH), with wavelength dependent dispersion parameters, shown in **Figure S4**. In the case of spectral range of interest, from 400 nm to 700 nm, the real part of refractive index steadily decreases from 5.1 to 4.3, but sustains the high refractive index and compared to the surrounding materials, air ($n = 1$) and glass ($n = 1.46$), high contrast in the system. In the same spectral range, the imaginary part of the refractive index, which is representing the losses in the material, drops significantly from 2.2 to approximately 0.2. Nevertheless, amorphous Si is expected to be lossy in the whole spectral range of visible light.

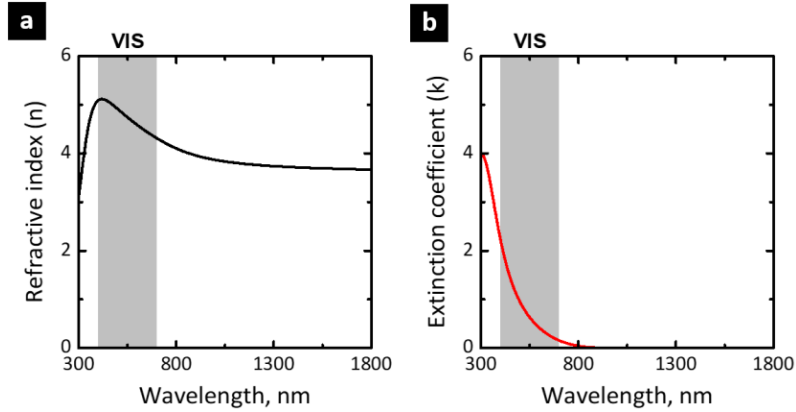

**Figure S4.** Dispersion parameters of amorphous silicon used in simulations and experiments. **(a)** Real part of the refractive index as a function of wavelength. **(b)** Imaginary part of the refractive index, extinction coefficient as a function of wavelength. Gray area depicts visible spectral range from 400 nm to 700 nm.

**S5. Angle Tolerance Analysis of Optimized RGB filters**

In the paper we presented and discussed the angle tolerance of the amorphous Si nanostructure-based RGB filters. The following filters were considered: R (diameter – 80 nm, period – 250 nm, height – 175 nm), G (diameter – 150 nm, period – 250 nm, height – 175 nm), B (diameter – 80 nm, period – 250 nm, height – 175 nm). In addition to the unpolarized illumination (both transverse electric (TE) and transverse magnetic (TM) modes), here were present results on the filter response via color quality of filters illuminated by a linearly polarized light. As the filters are relatively robust in case of TM polarization, see **Figure 5.1**(a), to understand the spectral dependence on angle of incidence, we examine the case of TE polarization, presented in **Figure 5.1**(b).

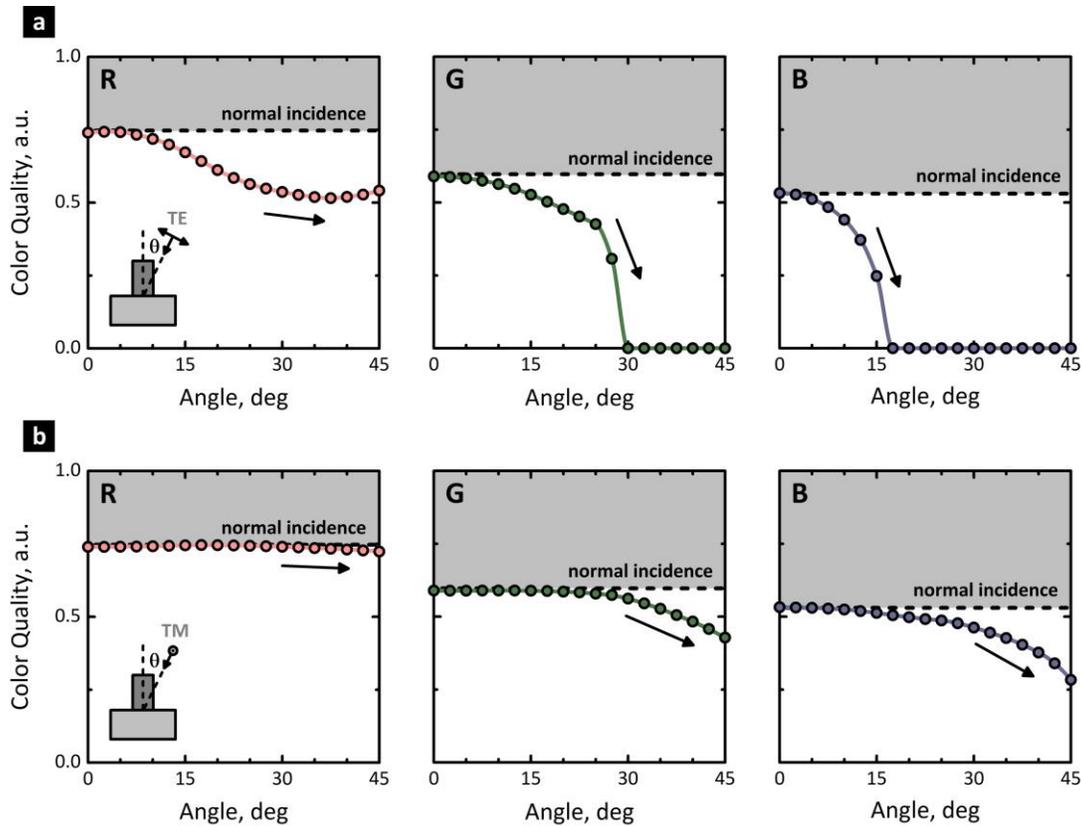

**Figure S5.1.** Color quality value change by the change of angle of incidence for red (left), green (middle) and blue (right) filters, compared to the normal incidence case (dashed line). **(a)** TE polarization. **(b)** TM polarization.

In **Figure 5.2**, we take a look at the E fields at cross-section of the nanostructures optimized for the R, G and B and illuminated with corresponding wavelengths: 650 nm, 550 nm and 450 nm. In contrast to normally incident light, at the angle of 15 degrees and 30 degrees we observe change in the electric field distribution and electric field lines inside of the nanostructure. The electric field distribution naturally changes as the structures are of a relatively small height (subwavelength) and by the increase of the angle light impinges at the different-size cross-section of the nanostructure. The coupling between the neighboring nanostructures also seems apparent as the electric field stretches to the boundaries of the simulation domain – periodic unit cell. As can be also seen, especially in the case of G filter, the loop comprising of the electric field lines shifts from the center of the nanostructure, it also appears to be stronger, which can be attributed to the blue-shift of the resonance.

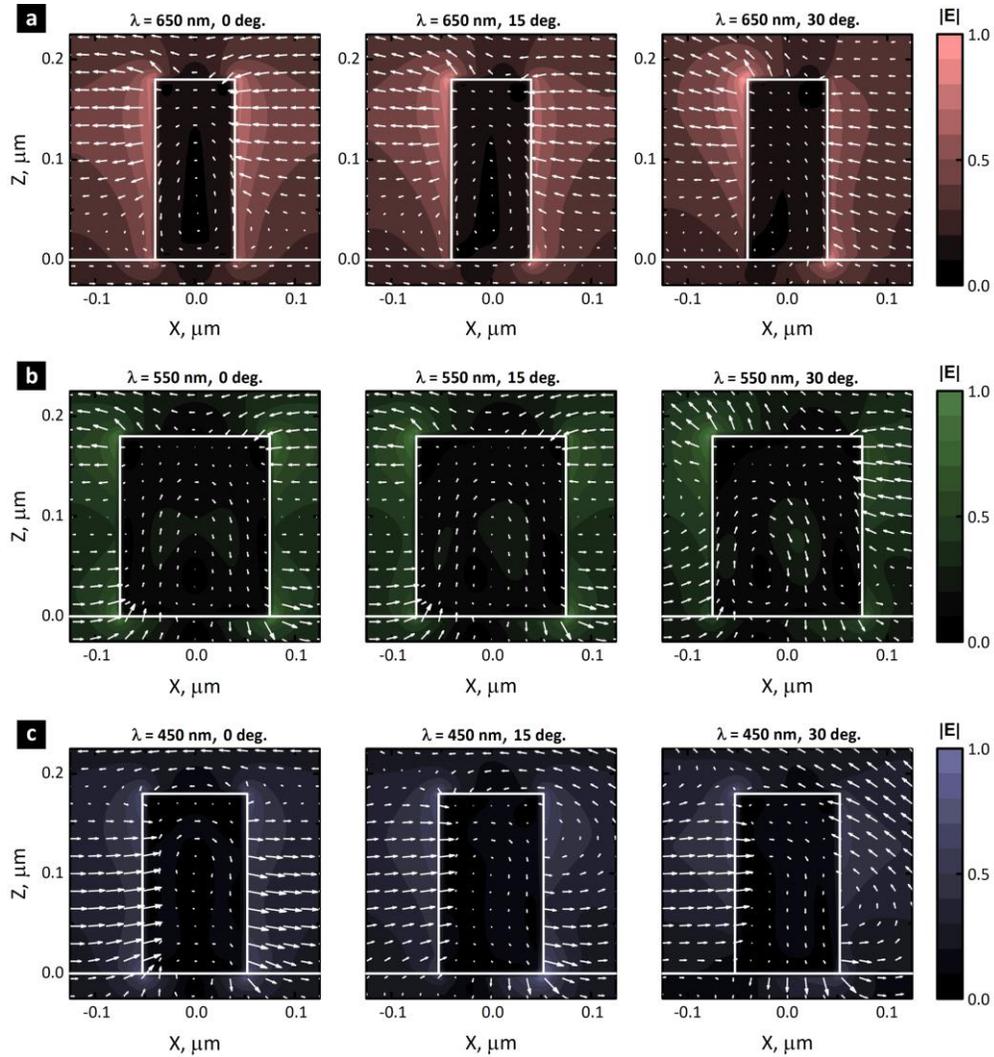

**Figure S5.2.** E field and E field lines at the cross-section of R, G, and B filters illuminated by TE polarized light at 0 degrees, 15 degrees, and 30 degrees of incidence angle (from left to right). **(a)** R filter illuminated by 650 nm wavelength. **(b)** G filter illuminated by 550 nm wavelength. **(c)** B filter illuminated by 450 nm wavelength. E field is normalized to the highest value, E field lines depicted by white arrows.

## S6. Spacer Thickness Influence on Color Quality

The spectral response of the sub-micrometer sized filter array depends on the spacer thickness, the thickness of the layer between the filters and the monitors, the CMOS sensor. In the emulation of the CMOS sensor the spacer thickness was fixed at 0.1 µm due to the suggested thickness in the work by another group[4]. However, as shown in **Figure 6**, our results support that 0.1 µm spacer is robust. In addition, judging by the color quality, the spacer thickness could be even increased up to 0.25 μm, which could potentially add more flexibility to the fabrication of the actual device.

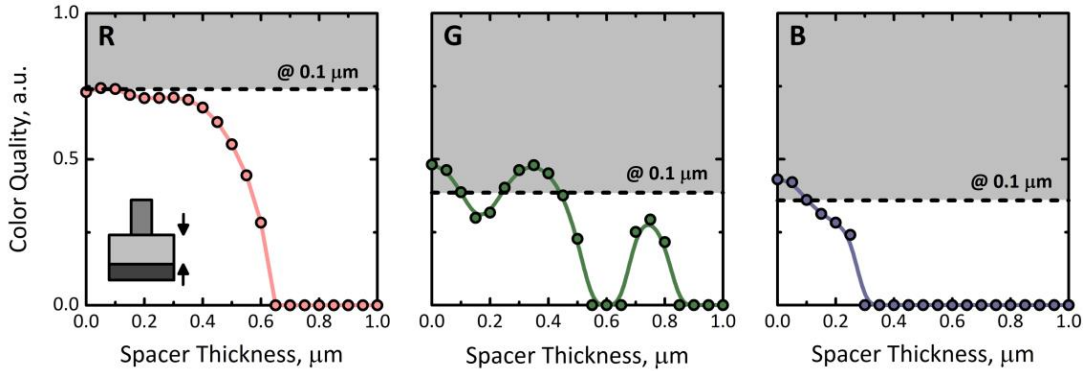

**Figure S6.** Color quality versus the thickness of the spacer. The thickness is varied from the interface of the silicon structures (no spacer) to the thickness of 1 μm.